\documentclass[12pt]{article}   
\usepackage[tablesfirst,notablist,nomarkers]{endfloat}
\usepackage{ol2} 

\begin{document}

\title{Ultra low frequency noise laser by locking to an
optical fiber delay line}


\author{Fabien K\'ef\'elian,$^{2,*}$ Haifeng Jiang,$^1$ Pierre Lemonde,$^1$  and Giorgio Santarelli$^{1}$}
\address{$^1$LNE-SYRTE, Observatoire de Paris and CNRS, \\ 61 avenue de l'Observatoire, 75014 Paris, France
}
\address{$^2$Laboratoire de Physique des Lasers, Universit\'e Paris 13 and CNRS,\\ 99 avenue Jean-Baptiste Cl\'ement, 93430 Villetaneuse, France}

\address{$^*$Corresponding author: fabien.kefelian at univ-paris13.fr}

\begin{abstract}
We report the frequency stabilization of an erbium-doped fiber
distributed-feedback laser using an all-fiber based Michelson
interferometer of large arm imbalance. The interferometer uses a 1
km SMF-28 optical fiber spool and an acousto optic modulator
allowing heterodyne detection. The frequency noise power spectral
density is reduced by more than 40 dB for Fourier frequencies
ranging from 1 Hz to 10 kHz, corresponding to a level well below 1
Hz$^2$/Hz over the whole range. It reaches 10$^{-2}$ Hz$^2$/Hz at
1 kHz. Between 40 Hz and 30 kHz, the frequency noise is shown to
be comparable to the one obtained by Pound-Drever-Hall locking to
a high finesse Fabry-Perot cavity. Locking to a fiber delay line
could consequently represent a reliable, simple and compact
alternative to cavity stabilization for short term linewidth
reduction.
\end{abstract}


 \noindent Very low frequency noise lasers are important tools
for many applications such as high-resolution spectroscopy,
optical atomic clock local oscillator, interferometric sensor
(including gravitational waves detection), and coherent optical
communications systems. Erbium-doped fiber distributed-feedback
lasers (DFB EDFL) near 1550 nm typically exhibit optical linewidth
in the range of 1-10 kHz and a frequency noise power spectral
density dominated by 1/$f$ component. Although narrower than diode
laser linewidth, these performances are still insufficient for
many of these applications.

Stabilization of laser frequency over very long term is achieved
by comparing it to an atomic or molecular absorption line
reference frequency. However this method generally does not
provide fast frequency noise reduction. The linewidth of lasers is
then usually reduced by locking to an ultra-stable optical cavity,
using the Pound-Drever-Hall method
\cite{Drever83}\cite{Ludlow07}\cite{Webster08}. It led to
fractional frequency instability lower than 10$^{-15}$ for 1 s
averaging times and subhertz linewidth \cite{Young99}. But this
scheme requires fine alignment of free space optical components,
tight polarization adjustment and spatial mode matching. Moreover,
high-finesse cavities are relatively expensive, bulky and fragile
devices. Finally to avoid air index fluctuations and improve the
thermal control and stability, the cavity has to be housed into a
high vacuum enclosure with thermal radiation shielding.

Another approach of the frequency noise reduction with a length
etalon is to use a 2 arm (Michelson or Mach-Zehnder)
interferometer to measure the frequency fluctuations during a
fixed time delay \cite{Chen88}\cite{Cranch02}. This method
requires a relatively large arm imbalance to obtain a sufficient
frequency discriminator sensitivity. Indeed, to have with a
Michelson interferometer the quality factor of a 10 cm Fabry-Perot
cavity of finesse 330000 the arm imbalance has to be as long as 10
km. Optical fiber is an excellent material to achieve such a large
path imbalance. This can lead to a largely lighter, more compact
and cheaper device than the low thermal expansion (ULE) high
finesse Fabry-Perot cavity. In addition a frequency shifter can be
inserted in one arm of the interferometer and consequently the
Pound-Drever-Hall system can be replaced by a simple heterodyne
detection scheme as detailed below. Thanks to the very small
effective free spectral range associated to a large arm imbalance
and to the continuous tunability of the system via a phase shift
on one of the interferometer arms, the frequency noise reduction
can operate at any well controlled laser frequency, as opposed to
Fabry-Perot cavity locking.

For Fourier frequencies where loop gain is high, the stabilized
laser relative frequency noise $\delta\nu(f)/\nu_0$ equals the
etalon optical length relative fluctuations $\delta L(f)/L_0$
which consequently has to be as low as possible. Ultra-stable
Fabry-Perot cavities use a low-thermal expansion material spacer
and have to be well protected from environmental perturbations
using vacuum enclosures and seismic isolation provided by modern
anti-vibration platforms. Their stability is ultimately limited by
fundamental thermodynamics fluctuations in mirrors substrates and
coatings \cite{Numata04}. The optical length of fibers is also
sensitive to acoustical, mechanical and thermal perturbations.
These perturbations are generally very low for Fourier frequencies
above 1 kHz but can significantly degrade the frequency noise of
the locked laser at low frequencies. The optical fiber must
consequently be placed into an enclosure to isolate it from
external fluctuations. If noise sources are spatially
uncorrelated, $\delta L(f)/L_0$ scales as $1/L_0^{1/2}$ and then
decreases with longer fiber. Finally, refractive index
thermodynamic fluctuations in the fiber set a fundamental limit on
the stabilized laser frequency noise \cite{Wanser92}. This limit
can be push down by extending the fiber length. However, this will
lower the cut-off and blind frequencies of the discriminator
transfer function. This can be partially overcome by designing
sophisticated loop filter controllers \cite{Sheard06}.

The first frequency stabilization experiment of a laser onto a
fiber spool used a Mach-Zehnder interferometer (MZI), with phase
modulation into one arm, to stabilize a He-Ne laser. Corrections
were applied via a PZT stretcher. It led to a 5 kHz linewidth on 1
s \cite{Chen88}. Cranch stabilized a DFB EDFL onto a 100 m path
imbalance MZI using homodyne electronics and PZT actuator,
reaching about 2 Hz$^2$/Hz at 1 kHz \cite{Cranch02}. The method
was also applied to diode lasers with all-fibered interferometers
using homodyne detection leading to 10$^2$ Hz$^2$/Hz at 10 kHz
\cite{Cliche06}. Extension of the control bandwidth using a
sophisticated controller design was achieved with a 10 km arm
imbalance Michelson interferometer \cite{Sheard06}. Recently,
frequency noise reduction of a DFB EDFL was reported to reach 4
Hz$^2$/Hz at 80 Hz \cite{Takahashi08}. The set-up is based on a
110 m arm imbalance all-fibered Michelson interferometer in vacuum
with advanced seismic isolation and homodyne detection.

\begin{figure}[htb]
\centerline{\includegraphics[width=8cm]{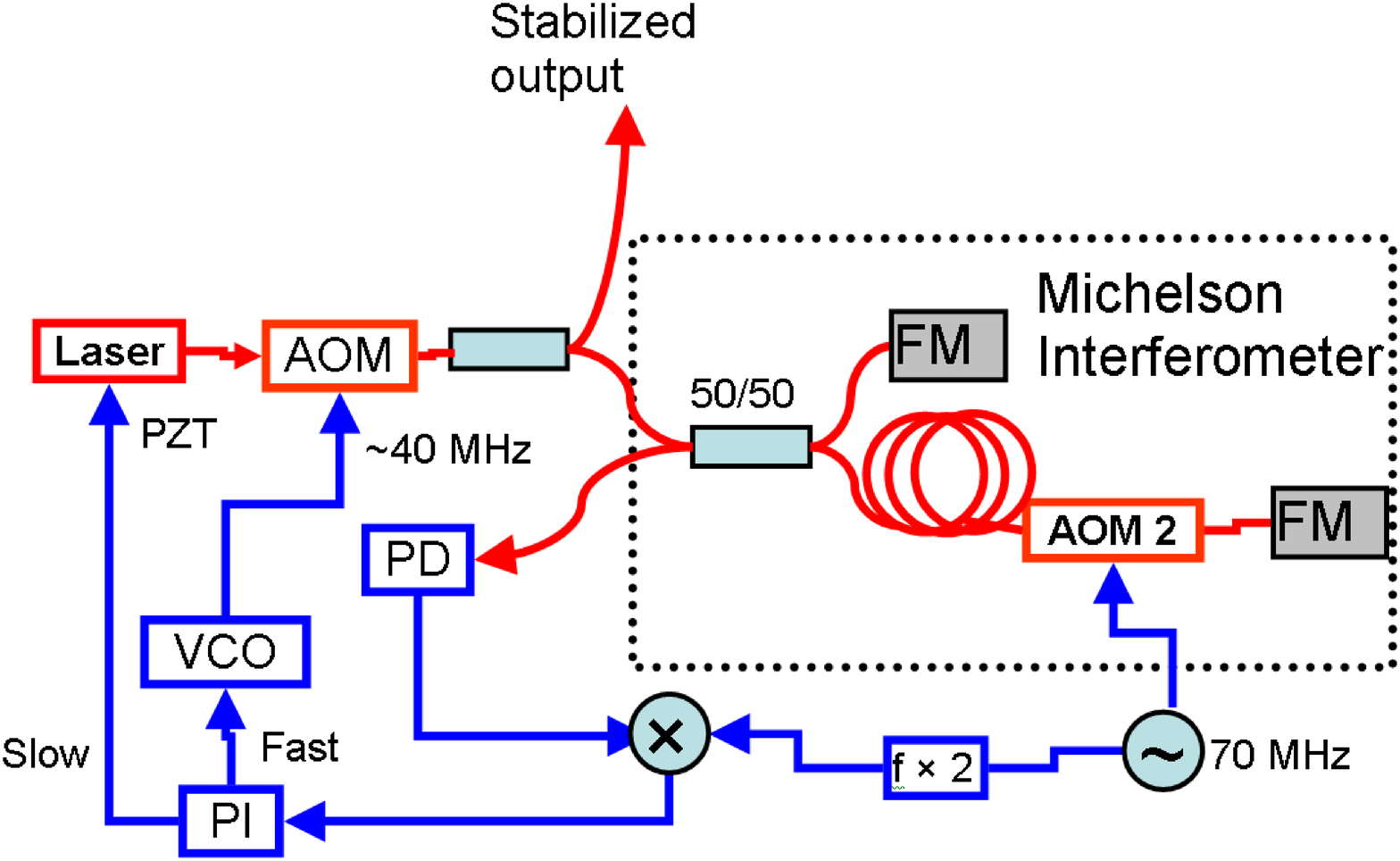}}
\caption{Scheme of the laser frequency noise reduction system;
AOM: acousto-optic modulator; PD: photodiode; VCO: voltage
controlled oscillator; PI: proportional-integrator filter; FM:
Faraday mirror} \label{fig:schema}
\end{figure}

We present here a system using an all-fibered 2 km imbalance
Michelson interferometer with heterodyne detection. Figure
\ref{fig:schema} shows the scheme of the laser frequency
stabilization system. The main component of the system is the
frequency-shifted fibered Michelson interferometer. The input
optical wave is split between the two arms by a 50/50 fiber
coupler. The first arm of the coupler is directly connected to a
Faraday mirror; the second arm of the coupler is connected to a 1
km spool of standard SMF-28 fiber followed by an acousto-optic
frequency shifter (AOM2) and a Faraday mirror (FM). The Faraday
rotator mirror guarantees that in a retracing fiber optic link the
output state of polarization is orthogonal to the entrance state,
consequently the two waves in the output port of the Michelson
interferometer have always the same state of polarization and lead
to a maximum beat-note signal amplitude without requiring any
polarization controller. The fiber spool is placed into a ring
shape aluminum box and the interferometer is placed inside a thick
aluminum box, recovered by a thermal isolating thermoplastic film,
which is set onto a compact seismic [310x310x170 mm] vibration
isolation platform. The whole experiment is covered by an acoustic
isolation box of volume $\approx$0.125 m$^3$.

\begin{figure}[htb]
\centerline{\includegraphics[width=8cm]{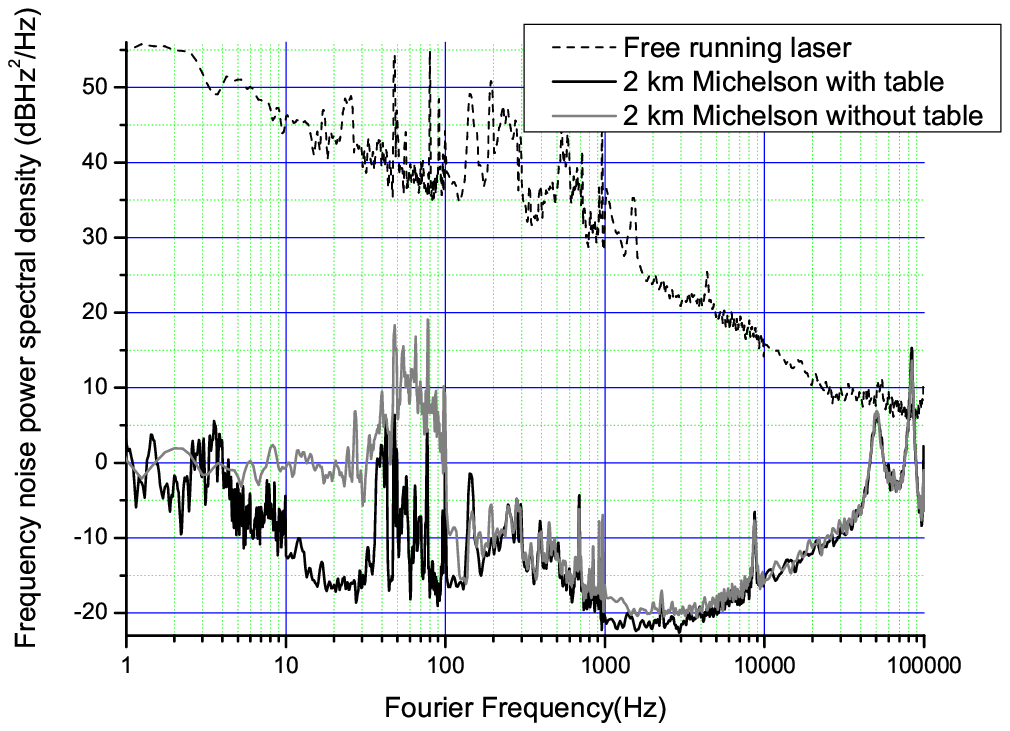}}
\caption{Frequency noise power spectral density versus Fourier
frequency of the free-running laser (dash line) and laser
stabilized on a 2 km imbalance Michelson interferometer with (dark
line) and without (gray line) passive anti-vibration table}
\label{fig:graph1}
\end{figure}

The frequency-shift fibered Michelson interferometer acts as an
optical frequency ($\nu_{opt}$) to RF phase ($\phi_{err}$)
converter with transfer function
$\phi_{err}(f)/\nu_{opt}(f)=H_{Mich}(f)=[1-e^{-i2\pi f\tau}]/if$
(rad/Hz), where $\tau$ is the fiber double-pass delay time and $f$
the Fourier frequency. For $f<<1/\tau$,
$H_{Mich}(f)\approx2\pi\tau$. The interferometer is seeded with
200 $\mu$W and the total double pass optical power losses in the
long arm containing the fibered acousto-optic modulator at 70 MHz
is about 6 dB. The optical power photodetected at the output port
of the interferometer then contains an RF carrier at $2f_{AOM}$
phase modulated by $\phi_{err}$ which is downconverted by an FR
mixer driven by the frequency doubled output of a low noise
reference oscillator at 70 MHz. This provides a low frequency
error signal proportional to $\phi_{err}+\Delta\theta_{RF}$, where
$\Delta\theta_{RF}$ is the local oscillator phase shift. The error
signal is amplified, filtered and converted into optical frequency
correction using AOM operating at 40 MHz which is driven by a
high modulation bandwidth voltage controlled oscillator for fast
correction and a piezoelectric element controlling the fiber laser
cavity length for drift compensation. The correction bandwidth
($\sim$100 kHz) is limited by the round trip delay in the fiber
interferometer. The laser source is a single-longitudinal-mode
Er$^{3+}$-doped fiber Bragg grating laser with an emission
wavelength of ~1542 nm and a maximum output power of 100 mW. The
free running laser frequency noise is dominated by a flicker
component with 10$^4$ Hz$^2$/Hz at 100 Hz as shown on Fig.
\ref{fig:graph1}.

\begin{figure}[htb]
\centerline{\includegraphics[width=8cm]{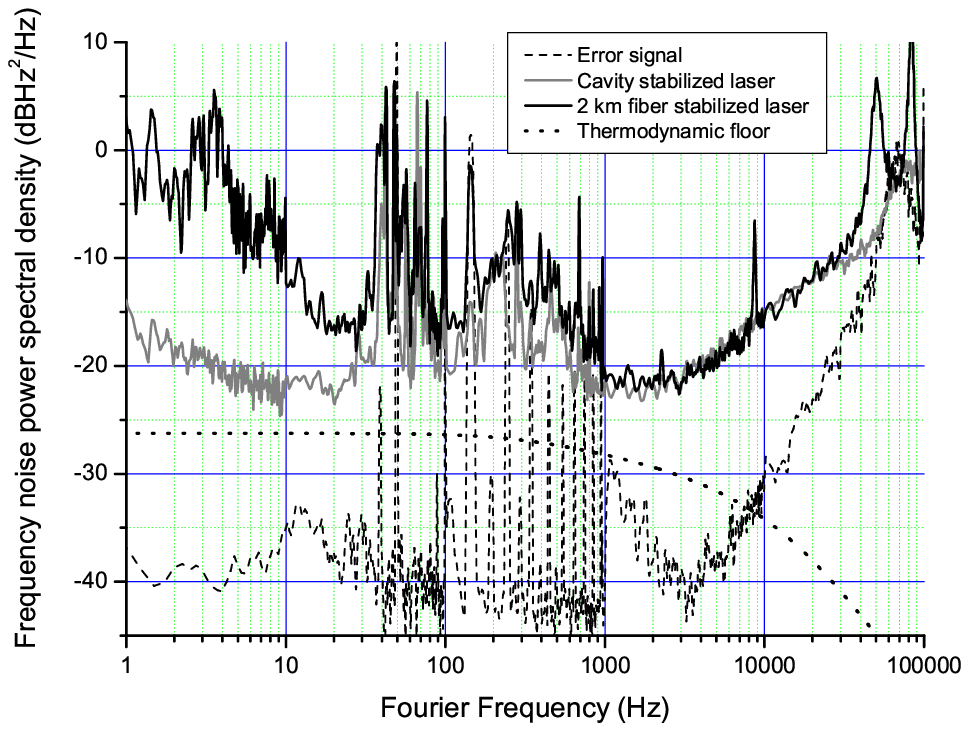}}
\caption{Frequency noise power spectral density versus Fourier
frequency of the laser stabilized on a 2 km imbalance Michelson
interferometer with anti-vibration table (dark line), reference
laser (gray line), error signal converted into frequency noise
(dash line) and thermodynamic noise floor (dot line).}
\label{fig:graph3}
\end{figure}

The frequency noise power spectral density (PSD) of the fiber
stabilized laser is measured by comparison with a high-finesse
Fabry-Perot cavity stabilized laser described in \cite{Jiang08}.
The radio frequency beat-note signal is down converted to 700 kHz
by a low phase noise synthesizer, then frequency-to-voltage
converted and analyzed using a fast Fourier transform analyzer
after removing a linear drift of the order of 1 kHz/s. Results are
shown on Fig. \ref{fig:graph1}. With the anti-vibration platform,
the frequency noise reduction is larger than 40 dB between 1 Hz
and 10 kHz and the frequency noise PSD is, notwithstanding several
peaks, below 1 Hz$^2$/Hz in the same range. Effect of the
anti-vibration table is notable between 5 Hz and 100 Hz, however
even without table the noise reduction is better than 30 dB. On
Fig. \ref{fig:graph3} is plotted the measured frequency noise of
the fiber stabilized laser together with the cavity stabilized
reference laser frequency noise. For Fourier frequencies between
40 Hz and 30 kHz the measurement is limited by the reference laser
frequency noise and is therefore an upper limit of the fiber
stabilized laser noise. A lower limit is also plotted in Fig. 3,
set by the thermodynamic noise derived from \cite{Wanser92} up to
10 kHz and the in-loop noise derived from the error signal
fluctuations above that frequency. Below 40 Hz the measured
frequency noise is dominated by the reference fiber optical length
fluctuations.

The optical spectrum was numerically derived by fast Fourier
transform of the auto-correlation function computed from the
frequency noise PSD measured from 100 mHz to 100 kHz. A
measurement time was taken into account corresponding to a first
order filter of the frequency noise PSD. The 3 dB linewidth of the
optical spectrum is 8 Hz for 1 s measurement time. When the
isolation platform is disactivated the linewidth rises to 17 Hz
for the same measurement time. For comparison the linewidth of the
free running laser is 8 kHz.

In conclusion, we have demonstrated that an all-fibered
frequency-shift Michelson interferometer can be used to reduce the
frequency noise of a fiber laser to a level comparable to a
high-finesse Fabry-Perot stabilized laser for Fourier frequencies
ranging from 40 Hz to 30 kHz. This is several orders of magnitude
better than previous results of laser stabilization using fiber
delay lines. This improvement is most likely due to the use of RF
heterodyne detection at a frequency where technical issues like
laser intensity noise and detection noise are totally negligible.
This method now constitutes an interesting alternative to cavity
locking for applications where this frequency range is relevant.
It has the great advantage to provide a fibered system without any
optical alignment or polarization adjustment. It is therefore
intrinsically more compact, light and flexible than cavity-based
systems. For Fourier frequencies below 40 Hz, our system is
presumably limited by thermal fluctuations and mechanical
vibrations which can certainly be improved. The interferometer
could for instance be installed in a temperature stabilized vacuum
tank with several thermal shields. In addition a reference fiber
of lower thermal sensitivity could be used like specifically
designed photonic crystal fibers \cite{Dangui05} or liquid crystal
polymer coatings \cite{Naito01}. The ultimate limits of this
system are still an open question.



\end{document}